# Spherically restricted motion of a charge in the field of a magnetic dipole

Emilio Cortés[1,3] and D Cortés-Poza[2]
[1]*Departamento de Física, Universidad Autónoma Metropolitana Iztapalapa,*
*P.O. Box 55-534, México D.F. 09340, México.*
[2] *Center for Intelligent Machines, McGill University, Montreal, Canada.*
[3]*Member of SNI, México.*

E-mail: emil@xanum.uam.mx

**Abstract.** We study the restricted motion of an electric charge in a spherical surface in the field of a magnetic dipole. This is the classical non-relativistic Stöermer problem within a sphere, with the dipole in its centre. We start from a Lagrangian approach which allows us to analyze the dynamical properties of the system, such as the role of a velocity dependent potential, the symmetries and the conservation properties. We derive the Hamilton equations of motion and observe that in this restricted case the equations can be reduced to a quadrature. From the Hamiltonian function we find for the polar angle an equivalent one-dimensional system of a particle in the presence of an effective potential. This equivalent potential function, which is a double well potential, allows us to get a clear description of this dynamical problem. We are able to find closed horizontal trajectories, as well as their period. Depending on initial conditions, we can find also some bands covered by non-periodic trajectories, as well as the conditions for the presence of loops. Then we obtain by means of numerical integration different plots of the trajectories in three dimensional graphs in the sphere. This restricted case of the Stoermer problem, which is formally integrable, is still a nonlinear problem with a complex and interesting dynamics and we believe that it can offer the student a better grasp of the subject than the general three dimensional case.

## 1. Introduction

The motion of a charged particle in a magnetic field has been of interest since the discovery of cosmic rays, at the beginning of the twentieth century. This problem was studied extensively by Carl Stöermer [1] for many years, and it is now called Stöermer problem. As the cosmic rays are composed of energetic subatomic particles, the particle dynamics belongs to the relativistic domain. See an early review [2]. The relevance of the system was the modeling of the Aurora Borealis phenomenon and later, in the middle of the last century, it received an impulse with the discovery of the van Allen belts. During the first half of the last century, physicists and mathematicians considering Earth's magnetic field in the first approximation as a bar magnet, were interested in solving the problem of describing the dynamics of a charge in a magnetic dipole field. Henri Poincaré [3] could solve the motion of charged particles near an isolated magnetic pole, showing that they spiraled around field lines and they were repelled from regions of strong field (the poles).

Many analytical and numerical models have been developed to study this problem. Special interest has been focused in the study of trapped orbits [1]- [4], among the many different classes of orbits that depend on the initial conditions. The general problem has three degrees of freedom and as the magnetic field of the dipole has an axis of symmetry and the energy is conserved, two constants of motion are readily identified. After decades of searching for an additional integral of motion, it was concluded that the problem is not integrable, actually it has been shown to be chaotic [5],[6].

In the present work we tackled the restricted case of the classical problem of a charge constrained to a spherical surface, interacting with the magnetic dipole in the centre of the sphere. Although the model can no longer be applied to the study of the cosmic radiation and the Earth's field, we believe that it may illustrate very clearly the features of this complex magnetic field through a rather simple physical system. We have not find in the literature the analysis of this restricted Stöermer problem.

The work is organized in the following way. In section 2 we establish the problem through the Lagrangian approach and discuss some important properties of the dynamical system. In section 3 we derive the Hamilton equations and show that they are formally integrable. In section 4 we analyse the dynamics in the polar angle variable and find the equivalence of the system with a one dimensional problem with an effective double well potential. We discuss different properties of the paths according to this framework. From a numerical integration of the equations of motion, we obtain 3D plots of different paths to compare with the analytical results. In section 5 we study the critical points of the equations of motion for $\dot{\theta}$ and $\dot{\varphi}$ and discuss the turning points and the presence of loops in the spherical surface, as well as other properties of the path with respect to the azimuthal motion. Finally in section 6 we present some conclusions.

## 2. Dynamical system

We introduce spherical coordinates, $R, \theta, \varphi$, to determine the position of the particle at the point $P$, we set $R$ constant, and we just have two coordinates, the polar and azimuthal angles, see figure 1. We set the magnetic dipole in the centre of the sphere with its magnetic dipole moment $\mathbf{m}$ pointing downwards. We define $M$ as the mass of the particle, the axes $x$ and $y$ are located in the equatorial plane and $z$ is the vertical or symmetry axis. As we know, with this selection we have the south magnetic pole in the direction of the upper hemisphere. This is the case of the magnetic Earth's field. The magnetic vector potential is then expressed as

$$\mathbf{A} = \frac{\mu_0 \mathbf{m} \times \mathbf{a_R}}{4\pi R^2}, \tag{1}$$

where $\mu_0$ is the magnetic suceptibility of vacum, and $\mathbf{a_R}$ is a unit vector in the radial direction.

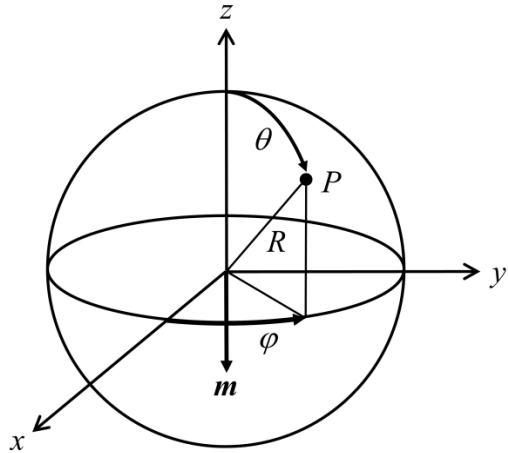

**Figure 1.** Spherical surface where the charge is constrained. The magnetic dipole is located at the centre of the sphere, with its dipole moment pointing along the negative $z$ axis. We will use $(R, \theta, \varphi)$ as the spherical coordinates, and $(x, y, z)$ as the rectangular coordinates of a point $P$.

From figure 1 we can write the vector $\mathbf{A}$ in the form

$$\mathbf{A} = \frac{-\mu_0 m \sin\theta \, \mathbf{a}_\varphi}{4\pi R^2}, \tag{2}$$

where $\mathbf{a}_\varphi$ is a unit vector in the azimuthal direction.

Now, instead of starting with the Lorentz force, we prefer to follow the Lagrangian formalism due to the fact that from this, we are able not only to derive more directly the equations of motion of the charge, but we can also obtain some important dynamical features like the symmetries and the conserved quantities.

The Lagrangian of the system has the form

$$L = K - U, \tag{3}$$

where $K$ is the kinetic energy of the particle and $U$ is a velocity dependent potential which it is shown [7] to be

$$U = -q\mathbf{v} \cdot \mathbf{A}, \tag{4}$$

where $\mathbf{v}$ is the velocity vector and $q$ is the charge of the particle. In spherical coordinates we have

$$\mathbf{v} = R\dot{\theta}\,\mathbf{a}_\theta + R\sin\theta\,\dot{\varphi}\,\mathbf{a}_\varphi, \tag{5}$$

where we use the unit vectors $\mathbf{a}_\theta$ and $\mathbf{a}_\varphi$.

Substituting eqs. (2) and (5) in eq. (4) we have

$$U = k\dot{\varphi}\sin^2\theta, \tag{6}$$

where we are defining

$$k = \frac{q\mu_0 m}{4\pi R}. \tag{7}$$

From eq. (5) the kinetic energy is written as

$$K = \frac{1}{2}MR^2(\dot{\theta}^2 + \dot{\varphi}^2\sin^2\theta). \tag{8}$$

Then the Lagrangian eq. (3) is written as

$$L = \frac{1}{2}MR^2(\dot{\theta}^2 + \dot{\varphi}^2\sin^2\theta) - k\dot{\varphi}\sin^2\theta, \tag{9}$$

The canonical conjugate momenta are defined as

$$p_\theta = \frac{\partial L}{\partial \dot{\theta}} = MR^2\dot{\theta}, \tag{10}$$

$$p_\varphi = \frac{\partial L}{\partial \dot{\varphi}} = \left[MR^2\dot{\varphi} - k\right]\sin^2\theta. \tag{11}$$

We observe that the term defined by

$$\eta \equiv MR^2\dot{\varphi}\sin^2\theta, \tag{12}$$

is the component of the angular momentum of the particle along the symmetry axis, $z$. We see here that as $\varphi$ is an ignorable coordinate due to the symmetry then $p_\varphi$, the conjugate momentum associated with this angle, is a constant of motion, but it is not an angular momentum; it has a term due to the magnetic field, that does not depend on a velocity.

Then we get our first conserved quantity,

$$p_\varphi = \eta - k\sin^2\theta = const. \tag{13}$$

The Hamiltonian of the system is defined in this case as

$$H = \dot{\theta}p_\theta + \dot{\varphi}p_\varphi - L. \tag{14}$$

From eqs. (10) and (11), first we write the function $H$ in terms of angles and velocities, after the cancellation of two terms we obtain

$$H = \frac{1}{2}MR^2(\dot{\theta}^2 + \dot{\varphi}^2\sin^2\theta). \tag{15}$$

and we find that $H$ is just the kinetic energy of the particle. We observe that the contribution of the magnetic field is implicit in the second term of this expression. As we know, the Hamiltonian here is another conserved quantity because $L$ does not depend explicitly on time. So this means that the kinetic energy is conserved, therefore the speed of the particle is also a constant. We recall that a static magnetic field does not produce work on a moving charge, although in this case it produces a torque. The velocity dependent potential given by eq. (4) does not contribute to the energy of the particle.

We identify two symmetries in the system, which means that the Lagrangian is invariant under translations in the time $t$ and in the azimuthal angle $\varphi$. See Noether's theorem [8]. These symmetries correspond to the two constants of motion, the Hamiltonian or kinetic energy and the generalized momentum $p_\varphi$. The angular momentum for arbitrary initial conditions is not conserved; we observe here that as the number degrees of freedom does not exceed the number of constants of motion, then the equations of motion are formally integrable. We point out that in the non-restricted Stormer's problem, where the radial distance is another variable, we have then three degrees of freedom but again the same two conserved quantities; in that case the dynamics happens to be chaotic [5], [6].

Now we can write the Hamiltonian again from eqs. (10) and (11) in terms of the coordinates and momenta and obtain

$$H = \frac{1}{2MR^2}\left[ p_\theta^2 + \frac{(p_\varphi + k \sin^2 \theta)^2}{\sin^2 \theta} \right] \qquad (16)$$

Then in the second term of the expression, eq.(16), we now see explicitly that the contribution of the magnetic field to the kinetic energy of the particle occurs in the azimuthal motion, and it is not a function of any velocity.

Using the integral of motion given by eq. (16) we can write

$$p_\theta = \pm \left( 2MR^2 H - \eta^2 / \sin^2 \theta \right)^{1/2}, \qquad (17)$$

Substituting in eq. (10) we obtain a quadrature

$$t = \pm MR^2 \int_{\theta_0}^{\theta} \left( 2MR^2 H - \left(\frac{\eta}{\sin \theta}\right)^2 \right)^{-1/2} d\theta. \qquad (18)$$

From the function $\theta(t)$ that could be obtained formally from the last equation, we would integrate eq. (11) to obtain the function $\varphi(t)$. The integral given by eq. (18) cannot be found in terms of analytic functions, so a numerical procedure would be necessary, but instead of this it will be more convenient to integrate the Hamilton equations numerically, which are going to be derived next.

## 3. Hamilton equations

From the Hamiltonian, eq. (16), and using eq. (12), first we have

$$\dot{\theta} = \frac{\partial H}{\partial p_\theta} = \frac{p_\theta}{MR^2}, \qquad (19)$$

$$\dot{\varphi} = \frac{\partial H}{\partial p_\varphi} = \frac{\eta}{MR^2 \sin^2 \theta}. \qquad (20)$$

These two equations could have already been written from eqs. (10) and (11). The other two equations are

$$\dot{p}_\theta = -\frac{\partial H}{\partial \theta} = \frac{\eta}{MR^2 \sin^2 \theta}[\eta \cot \theta - k \sin 2\theta], \qquad (21)$$

$$\dot{p}_\varphi = -\frac{\partial H}{\partial \varphi} = 0. \qquad (22)$$

The last equation was of course already given by eq. (13).

We point out that the two poles (north and south) are singular points, see eqs. (20) and (21). We assume that the polar angle is restricted to the open interval between $0$ and $\pi$, which is justified as long as the trajectory of the particle does not go directly into any of the poles. Actually, we see in equation (20) that $\dot{\varphi}$ grows indefinitely when the particle approaches the poles, so the azimuthal angular momentum also increases and this reduces the chances for the particle reaching these two points. As a matter of fact, as we will see in the next section, the particle is subject to a repulsive effect when it approaches the poles [3].

We see in the Hamilton equations, that eqs. (19), (20) and (16) can be integrated formally to give a function $\varphi(\theta)$:

$$\varphi(\theta) = \pm \int_{\theta_0}^{\theta} \frac{\eta \, d\theta}{\sin^2 \theta \sqrt{2MR^2 - \eta^2 \csc^2 \theta}}. \qquad (23)$$

This integral cannot be expressed in terms of known functions, so a numerical method would be required. This means that a full analytical solution of the Hamilton equations is not possible.

On the other hand, eqs. (19) and (21) constitute a coupled system of equations in the $\theta$ variable. If we divide equation (21) by equation (19) we can integrate an obtain an analytical expression for $p_\theta(\theta)$, we can write

$$p_\theta(\theta) = \pm \left[ p_{\theta 0}^2 + 2\int_{\theta_0}^{\theta} \eta(\eta \cot\theta - k\sin 2\theta)\csc^2\theta\, d\theta \right]^{1/2} \qquad (24)$$

$$= \pm \left[ p_{\theta 0}^2 + k^2(\cos^2\theta - \cos^2\theta_0) - p_\varphi^2(\cot^2\theta - \cot^2\theta_0) \right]^{1/2},$$

where $p_{\theta 0}$ stands for $p_\theta(0)$.

## 4. The phase space $(\theta, p_\theta)$ and the effective potential

*4.1 The phase space*

We see that the domain of this last function may exclude some intervals of angles, depending on the values of the parameters. In figure 2 we plot this function from 0 to $\pi$, with, $R=10$ and $k=0.5$, the initial values $p_\theta(0)=0$, $\theta(0)=\pi/9$, and for the constant we chose two values: $p_\varphi = 0.34k$, for which the curve exhibits two separated islands, and $p_\varphi = 0.36k$ for which one obtains a simple closed curve; all the values in arbitrary units. Here we point out that the initial values $p_\theta(0)$ and $\theta(0)$ determine the energy of the system, see eq. (16).

We clearly observe a symmetry with respect to the equator, $\theta = \dfrac{\pi}{2}$. This curve is similar to what it is obtained in a "bistable potential", whose phase space diagram can consist of one single closed curve or two closed curves, depending on the energy value. We point out here that the projection of the particle trajectory on the $(\theta, p_\theta)$ plane is always a closed path. The path in the spherical surface in general will be non-closed due to the azimuthal motion. In other words, in the three dimensional space $(\theta, p_\theta, \varphi)$ the plot of the trajectory will have a helicoidally shape which lies on the surface of a torus.

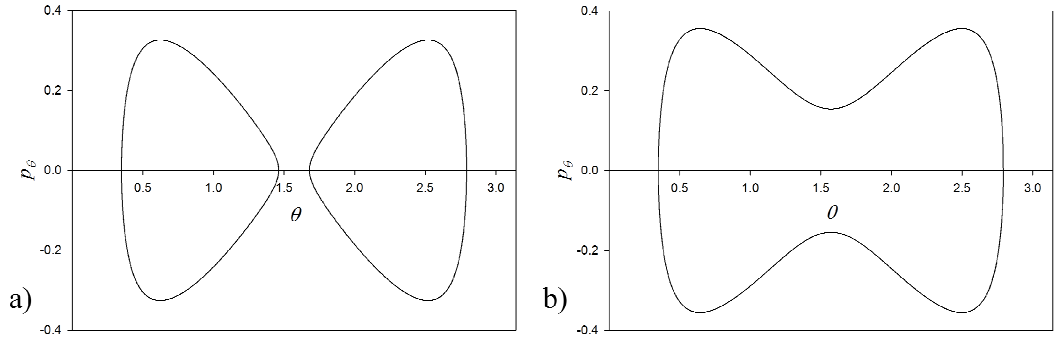

**Figure 2.** The phase space diagram of $p_\theta$ vs. $\theta$ for two different values for the constant $p_\varphi$. In (a) $p_\varphi = 0.34$ and in (b) $p_\varphi = 0.36$.

*4.2 The effective potential*

Let us write again eq. (16) for the Hamiltonian or kinetic energy

$$H = \frac{1}{2MR^2}\left[ p_\theta^2 + (p_\varphi + k\sin^2\theta)^2 \csc^2\theta \right]. \qquad (25)$$

We observe that the second term of this equation is the contribution of the azimuthal degree of freedom to the kinetic energy and, as we said, it does not depend on a velocity, but just on the polar coordinate. Then we can think in an equivalent one-dimensional problem, where the first term is the kinetic energy and the second one an "effective potential",

$$V_{\text{eff}}(\theta) = \frac{1}{2MR^2}(p_\varphi + k\sin^2\theta)^2 \csc^2\theta. \qquad (26)$$

Now we make a plot of this potential, see figure 3, with the following values for the parameters: $M = 2$, $R = 10$, $k = 0.5$ (in arbitrary units) and two different values for the constant: $p_\varphi = 0.2k$ and $p_\varphi = 0.95k$. It is effectively a symmetric bistable potential (or a single well), with its middle point in $\theta = \pi/2$ (the equator).

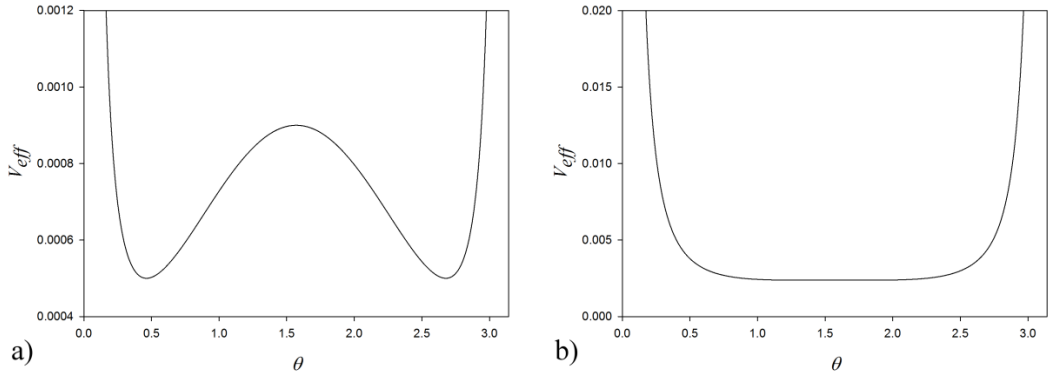

**Figure 3.** The effective potential of the equivalent one dimensional system in the $\theta$ variable, for two different values of the constant $p_\varphi$. In (a) $p_\varphi = 0.2k$ and in (b) $p_\varphi = 0.95k$.

There are some important features of our system that can be deduced from this potential. First we can appreciate a repulsive effect when the particle approaches the poles [3]. Then we can show from the expression (26) that for $|p_\varphi| \geq k$ the maximum in the potential disappears and we will have only one well between the two poles $\theta = 0$ and $\theta = \pi$. So we have as a condition for the bistability,

$$|p_\varphi| < k.  \qquad (27)$$

Now, in the presence of the double well, see figure 4, we have two stable equilibrium points $\theta_s$ and $\pi - \theta_s$, and one unstable equilibrium point at $\theta = \pi/2$. These three points correspond to horizontal paths ($\theta = const$). On each one of the wells (hemispheres) we observe the presence of two turning points, $\theta_1$ and $\theta_2$, where $p_\theta = 0$ and then the total energy is given by the value of $V_{eff}$ at this level. If the total energy is big enough, the particle will cross the barrier and it will travel periodically from one hemisphere to the other.

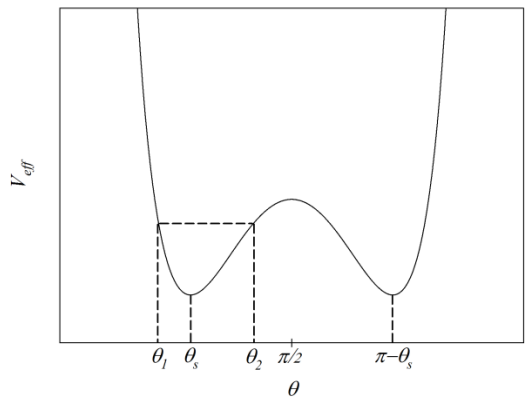

**Figure 4**. The effective potential showing its extreme values and turning points.

From eq. (26) we can take the derivative to identify its extreme values, where we use eq. (13)

$$V'_{eff} = \frac{\eta \csc \theta^3}{MR^2} \cos \theta (2k \sin^2 \theta - \eta) = 0. \qquad (28)$$

We have three possible solutions of this equation:

i) $\eta = 0$,

ii) $\eta = 2k \sin^2 \theta$,

iii) $\cos \theta = 0 \rightarrow \theta = \dfrac{\pi}{2}$.

(i) In the first solution, from eq. (12) we have $\dot{\varphi} = 0$ and, as we are on an extreme point of the potential then $\dot{\theta} = 0$, which means that the particle is at rest at any of the two stable equilibrium points. The corresponding angle $\theta_s$ is obtained from eq. (13)

$$\theta_s = \sin^{-1}\left[\dfrac{-p_\varphi}{k}\right]^{1/2}. \tag{29}$$

(We take the positive sign of the square root, because of the domain $0 < \theta < \pi$). Here we point out that the constant conjugate momentum $p_\varphi$ has to be negative as a necessary condition for the particle to be at rest. We notice that this conjugate momentum is not zero even when the particle is in a static equilibrium; this is due to the fact that from the definition of this quantity in eq. (11), it contains a term that depends just in the polar angle and not in a velocity.

(ii) In the second solution, from eqs. (12) and (13), we have:

$$p_\varphi = k \sin^2 \theta_s \tag{30}$$

$$\dot{\varphi} = \dfrac{\eta}{MR^2 \sin^2 \theta_s} = \dfrac{2k}{MR^2} \tag{31}$$

Here $\theta_s = \sin^{-1}\left[\dfrac{p_\varphi}{k}\right]^{1/2}$ (we take the positive sign due to the domain $0 < \theta < \pi$). The conjugate momentum $p_\varphi$, has to be positive as a necessary condition for this stationary value. This yields the two values of the minimum points in the bistable potential, see figure 4.

$$\theta_1 = \theta_s, \tag{32}$$
$$\theta_2 = \pi - \theta_s. \tag{33}$$

Then we have a horizontal closed trajectory at an angle $\theta_s$, see figure 5.

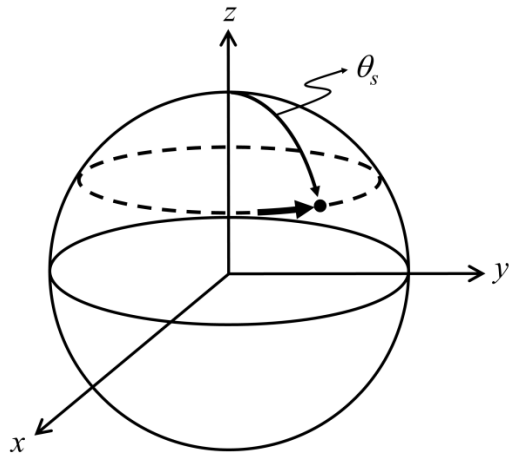

**Figure 5.** Horizontal closed path defined by $\theta = \theta_s$.

We see in eq. (31) that the angular velocity $\dot{\varphi}$ turns out to be independent of the polar angle and then the period for any of these horizontal paths is the same and has the value

$$T_h = \frac{\pi MR^2}{k}. \tag{34}$$

We note that the speed for these different horizontal paths depends on the angle $\theta_s$,

$$v = \dot{\varphi} R \sin \theta_s. \tag{35}$$

We remind here that the kinetic energy of the particle is a constant, once the parameters and initial values are given. So, the speed of the particle is also a constant for the given value of the angle $\theta_s$ in last equation. However, an interesting result is that the period for all these horizontal paths is the same for any polar angle different from $\pi/2$; we will examine this case in (iii).

(iii) The third solution ($\theta = \frac{\pi}{2}$) corresponds to an equatorial trajectory. From eq. (20) the azimuthal angular velocity is then

$$\dot{\varphi} = \frac{1}{MR^2}(p_\varphi + k) = const. \tag{36}$$

We observe here that as $-k < p_\varphi < k$ then the particle always travels to the east as long as we are in this "bistable regime".

We see that among all the horizontal closed trajectories, only those located in the equator ($\theta = \pi/2$), can have different angular velocities depending on the value of the constant $p_\varphi$. For any other value of the polar angle the angular velocity is given by the eq.(31).

### 4.3 Horizontal bands and the equatorial limit

Now, when $\theta(t)$ is changing, the path of the particle can be trapped in any of the two wells, if the condition of bistability is satisfied, which is $|p_\varphi| < k$, and the energy is smaller than the maximum of the potential. We chose $p_\theta(0) = 0$ (or some small value) and we set for the energy (given by the value of $\theta(0)$ according to eq. (26)), a value not bigger than the barrier height. What we have here is that the particle is trapped in a horizontal band defined by two values $\theta_1$ and $\theta_2$ which have the same potential energy, see the potential in figure 4 and the paths obtained numerically, figures 7. We point out that if $|p_\varphi| \geq k$, then the effective potential has just one well and the path will still be within a band, which crosses the equator. See figure 3b.

Now let the initial value of the polar angle, $\theta_0$ be in the open interval between 0 and $\pi/2$. We want to know the value of $p_\theta(\theta) \equiv p_{\theta 0}$ such that the particle will reach the equator tangentially, which means with $p_\theta = 0$, and from there will travel around the sphere between the angles $\theta_0$ and $\pi/2$. We are referring to this as the equatorial limit; in this case the particle will remain in the initial hemisphere. As the equator is in this case a turning point for the angle $\theta$, the energy of the particle would be

$$H = V_{eff}(\frac{\pi}{2}) = \frac{1}{2MR^2}(p_\varphi + k)^2 \tag{37}$$

But from eq. (16) at $t = 0$ we have

$$2MR^2 H = p_{\theta 0}^2 + (p_\varphi + k \sin^2 \theta_0)^2 / \sin^2 \theta_0. \tag{38}$$

Equating eqs.(37) and (38) and defining $c$ as

$$p_\varphi = ck \tag{39}$$

we obtain

$$p_{\theta 0}^2 = k^2 \left[(c+1)^2 - [(c+\sin^2 \theta_0)/\sin \theta_0]^2\right]. \tag{40}$$

Let us define the right side of the last expression as

$$\xi = (c+1)^2 - [(c+\sin^2 \theta_0)/\sin \theta_0]^2. \tag{41}$$

For $|c| \geq 1$, which is $|p_\varphi| \geq k$, we have $\xi < 0$ and then there is no solution of the eq. (40) in this case. This is in agreement with the condition for bistability given by eq.(27), so if this condition is not satisfied then there is no equatorial limit for the path.

But now, as we see in figure 6a, for $|c|<1$ we obtain $\xi >0$ within some subinterval of $0<\theta_0 \leq \pi/2$. In figure 6b we have the plot of the square root of the positive values of $\xi$ of figure 6a, multiplied by the constant $k$. In this plot we have the permitted values of $p_{\theta 0}$ for which the path will reach the equator tangentially, see eq. (40).

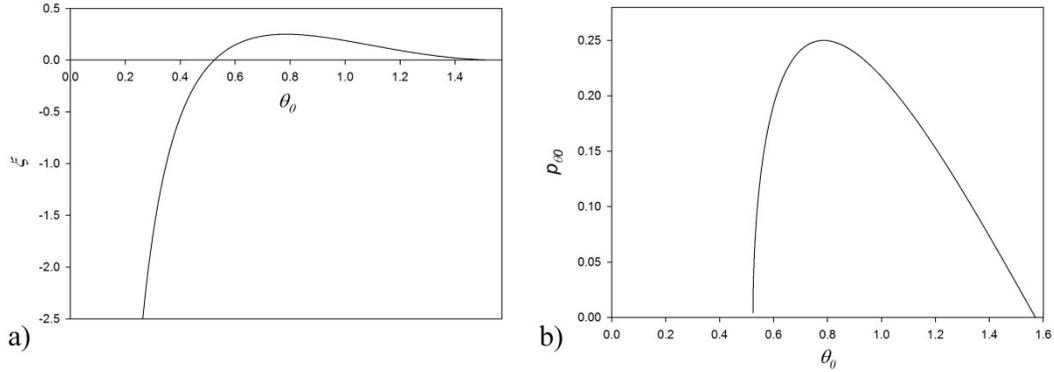

**Figure 6.** The domain of $p_{\theta 0}$ given by eq.(40) for the value $c=0.5$ (a) The plot of the function $\xi(\theta_0)$. (b) The plot of the interval of real values of $p_{\theta 0}$, as a function of $\theta_0$.

### 4.4 Spherical 3D plots

Now we can compare these last results to those obtained from the numerical integration of the Hamilton equations, eqs. (19)-(21), using a fourth order Runge-Kuta algorithm, which is going to be used for all the graphs in the rest of this work. We obtain some 3D graphs for the charge trajectory in different regimes. We are choosing for the mass and radius, $M=2$ and $R=10$, and $k=0.5$, in arbitrary units.

In figures 7 we see some paths trapped in a band in the northern hemisphere. In figures 8 we illustrate the case we call "equatorial limit", where we have the paths for different initial values of $p_\theta$, with the parameter $c=0.5$ used in figure 6; in all these paths the initial polar angle is $\theta_0 = 0.6$ rad. We are in the interval of possible values of $p_\theta$ of figure 6b. We present three different plots: in figure 8a we set $p_{\theta 0}=.1$, in figure 8b we set $p_{\theta 0}=.191725$ and in figure 8c we set $p_{\theta 0}=.25$. In figure 8b we clearly notice the equatorial limit of a path. In figure 8c the path crosses the equator and goes periodically from one hemisphere to the other.

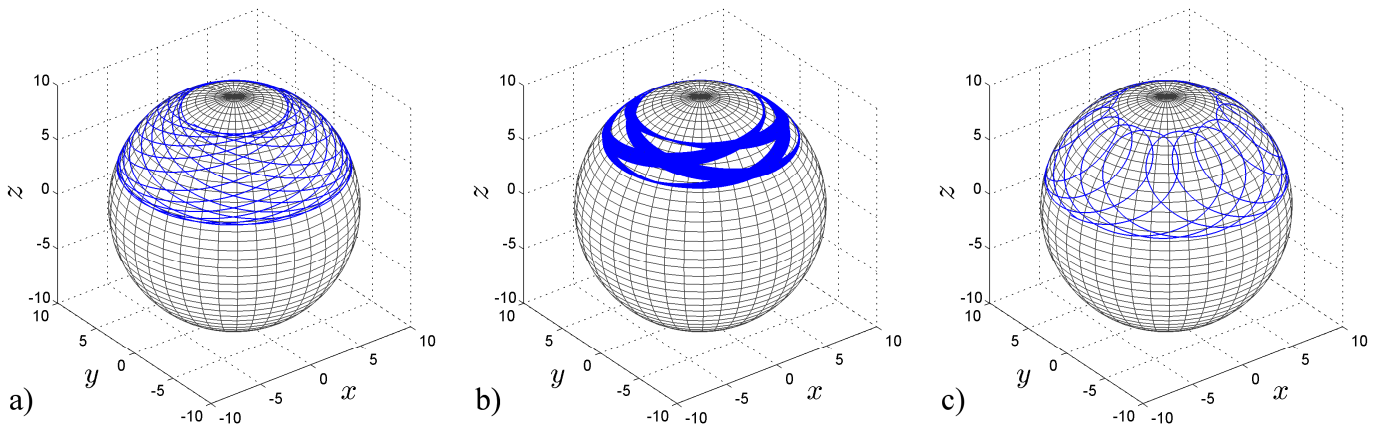

**Figure 7.** Three different paths in the sphere in a 3D plot. In all the cases the paths are trapped in a band between two polar angles in the northern hemisphere. In (a) we set $\theta_0 = \frac{\pi}{4}$, $p_{\theta 0}=0$ and $p_\varphi = 0.394$ in (b) we set $\theta_0 = \frac{\pi}{3}$, $p_{\theta 0}=0$ and $p_\varphi = 0.394$ in (c) we set $\theta_0 = \frac{75}{180}\pi$, $p_{\theta 0}=0$ and $p_\varphi = -0.394$. (In arbitrary units).

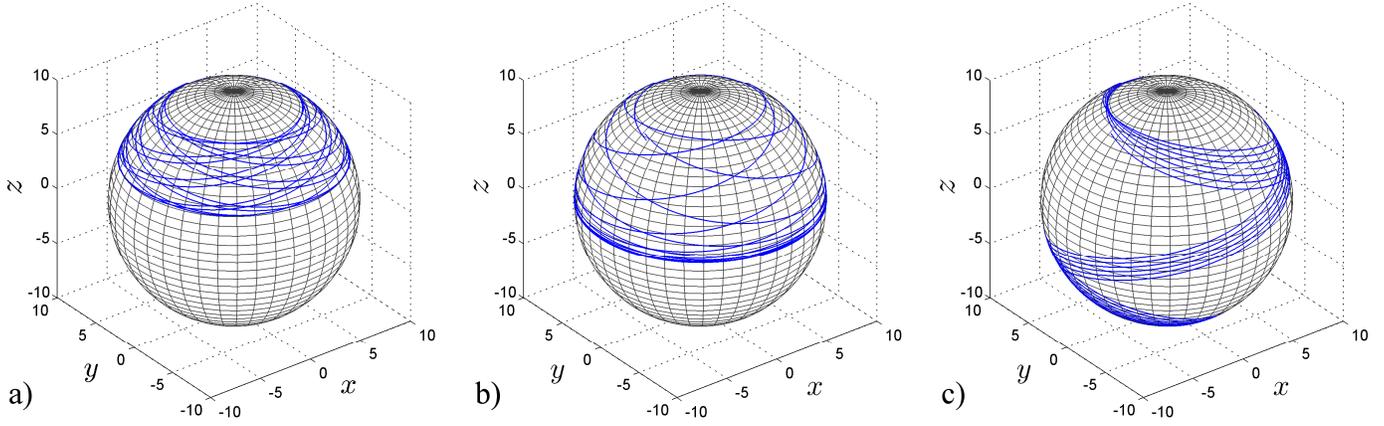

**Figure 8.** Three paths with different initial values of the moment $p_{\theta 0}$. In the three paths we set $\theta_0 = 0.6\,rad$, $p_\varphi = 0.5\,k$. In (a) $p_{\theta 0} = 0.1$ for which the particle remains in the northern hemisphere; in (b) $p_{\theta 0} = 0.191725$, for which the path reaches the equatorial limit and in (c) $p_{\theta 0} = 0.2525$, here the particle crosses and moves periodically between the two hemispheres.

## 4.5 Poincare section

We present a Poincare section of the path in the three dimensional space $(\theta, p_\theta, \varphi)$ by means of a projection of the trajectory into the $(\theta, p_\theta)$ plane by plotting the points in each period of the azimuthal angle, which is every time the azimuthal coordinate $\varphi$ crosses the positive $x$ axis. In figures 9 and 10 we have the Poincare section, as well as the 3D plots of the paths in the sphere, of two quasi-periodic paths, and in figure 11 we have the corresponding diagram for a non-periodic path.

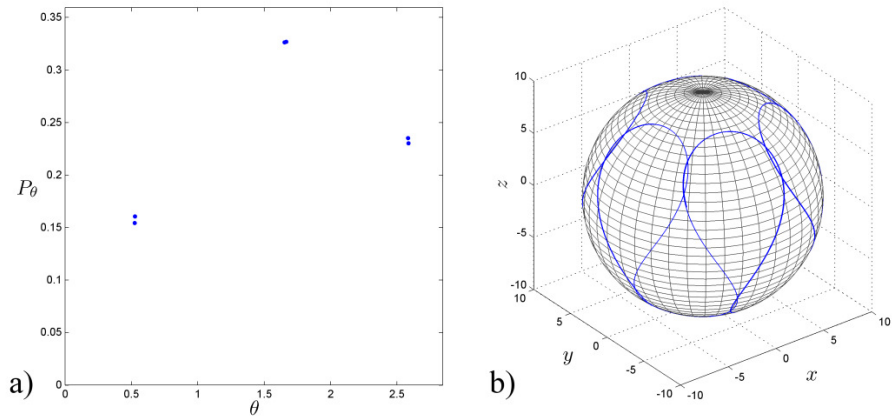

**Figure 9.** (a) Poincare section of the trajectory in the space $(\theta, p_\theta, \varphi)$. The points are sampled when $\varphi = 2n\pi$, $n = 1, 2, ...$, (which means $y = 0, x \geq 0$). (b) The corresponding 3D plot in the spherical surface. We are setting the value $p_\varphi = -0.6\,k$, which produces loops (see section 5), and the initial values $\theta_0 = \frac{\pi}{6}$ and $p_{\theta 0} = 0.151$ (in arbitrary units).

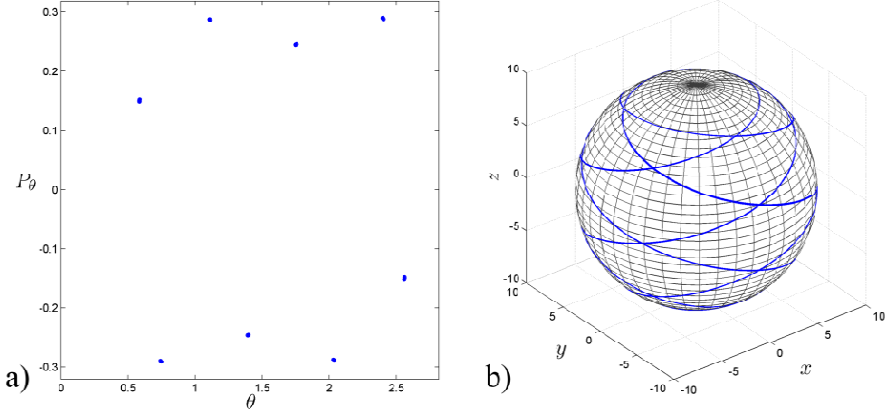

**Figure 10.** A similar graph to figure 9. Here we are setting the value $p_\varphi = 0.601k$, and the initial values $\theta_0 = \frac{\pi}{4}$ and $p_{\theta 0} = 0.3$ (in arbitrary units).

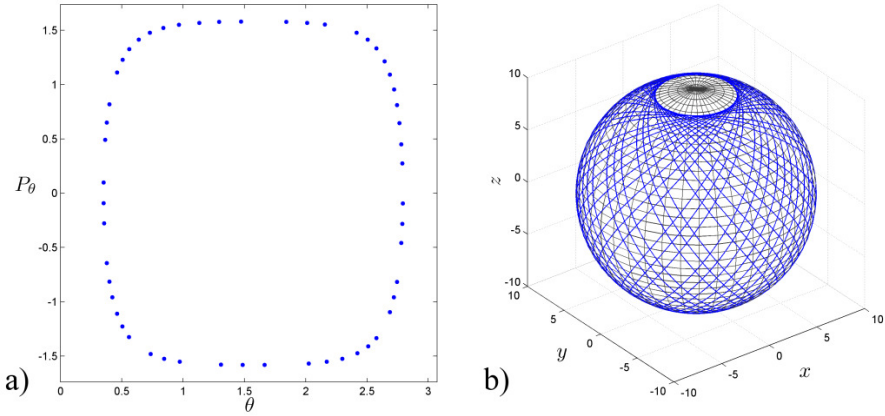

**Figure 11.** Similar graph to figures 9 and 10. Here we are setting the value $p_\varphi = 1.2k$, and the initial values $\theta_0 = \frac{\pi}{9}$ and $p_{\theta 0} = 0$ (in arbitrary units).

## 5. Critical points: turning points and loops

In the last section we have studied the behaviour of the dynamics of the polar angle coordinate, and we already described the critical points $\dot{\theta} = 0$, as the turning points in the effective potential plot. Now, we want to consider the dynamics of the azimuthal angle. We want to see what is the behaviour of the equation (20) in the critical points, where the angular velocity $\dot{\varphi}$ vanishes.

We see that if the azimuthal angular velocity is zero and if the two poles are excluded, then the azimuthal angular momentum is zero,

$$\eta = 0, \qquad (42)$$

which from eq. (13) we obtain

$$\sin\theta \equiv u = +\sqrt{-\frac{p_\varphi}{k}}. \qquad (43)$$

As the domain of the polar angle is $0 < \theta < \pi$, then we exclude the negative sign in the previous equation.

This result means that the condition $\dot{\varphi} = 0$ requires the condition

$$-k < p_\varphi < 0. \qquad (44)$$

In this angle, eq. (43), the trajectory of the particle changes the sign of the azimuthal velocity. The angle is the same as the minimum of the potential given in eq. (29), and it is located between the angles $\theta_1$ and $\theta_2$, which are the turning points, where $\dot{\theta} = 0$, see figure 4. As $\dot{\varphi}$ varies periodically with $\theta$, see eq. (11), then we have here the presence of loops in the trajectory. So we see that there is an out-of-phase synchronization between $\dot{\theta}$ and $\dot{\varphi}$; so this means that under this regime, given by the inequality (44), we will have the presence of loops in the $(\theta, \varphi)$ surface, as we can see in figure 12. We want to make a remark about these paths in the spherical surface. We could have in some paths the presence of "pseudo-loops", in those cases when the apparent loops (even being quite narrow) go around any one of the poles, see figure 8c; in these cases $\dot{\varphi}$ does not change its sign. In other words, the authentic loops do not enclose any pole of the sphere. We observe that all the paths of figure 8 are obtained for a positive value of $p_\varphi$ which is different to the condition (44).

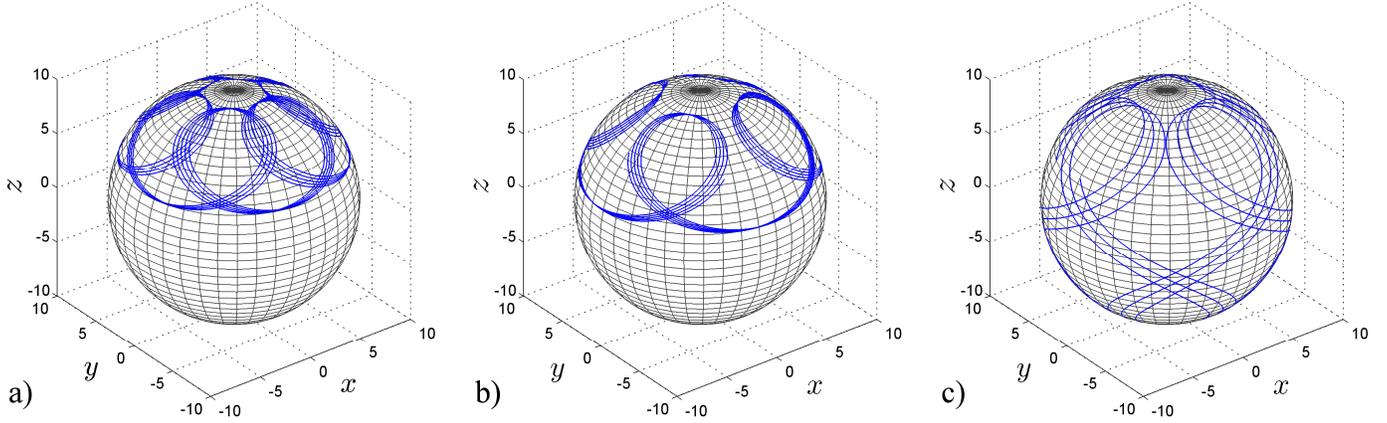

**Figure 12.** We have three different paths that, according to the condition given by eq. (44), exhibit the presence of loops in the spherical surface. In (a) we have $\theta_0 = \frac{\pi}{6}$, $p_\theta = 0.34$ and $p_\varphi = -.25\,k$. In (b) we have $\theta_0 = \frac{\pi}{4}$, $p_{\theta 0} = 0.3$ and $p_\varphi = -0.3\,k$. In (c) we have $\theta_0 = \frac{\pi}{3}$, $p_{\theta 0} = 0.3$ and $p_\varphi = -0.3\,k$.

Now, with respect to the direction of the azimuthal motion of the particle we can take the eqs. (13) and (20) and write for the angular velocity

$$\dot{\varphi} = \frac{p_\varphi + k \sin^2 \theta}{MR^2 \sin^2 \theta}. \tag{45}$$

From this equation we can classify different trajectories, according to the azimuthal direction of the motion,

a) If $p_\varphi \geq 0$ then $\dot{\varphi}$ is always positive (eastwards in Earth's coordinates).
b) If $p_\varphi < -k$ then $\dot{\varphi}$ is always negative (westwards in Earth's coordinates).
c) If $-k \leq p_\varphi < 0$ then $\dot{\varphi}$ can changes sign according to the variations of $\theta$. Here we have the presence of loops in the trajectory. See figures 12.

Finally, it is important to point out that in the presence of loops, the path could have a drift motion in either east or west directions, depending on the precise numerical value of the constant $p_\varphi$, and the path could be like those of the figures 12. But it is possible to find a critical value of $p_\varphi$ (given $\theta_0$ and $p_{\theta 0}$), for which there is no drift in the azimuthal motion and the path will be periodic with oscillations back and forth in the angle $\varphi$, see figure 13. We obtain this critical value of $p_\varphi$ by swiping its values from which we get a change between eastward to westward directions.

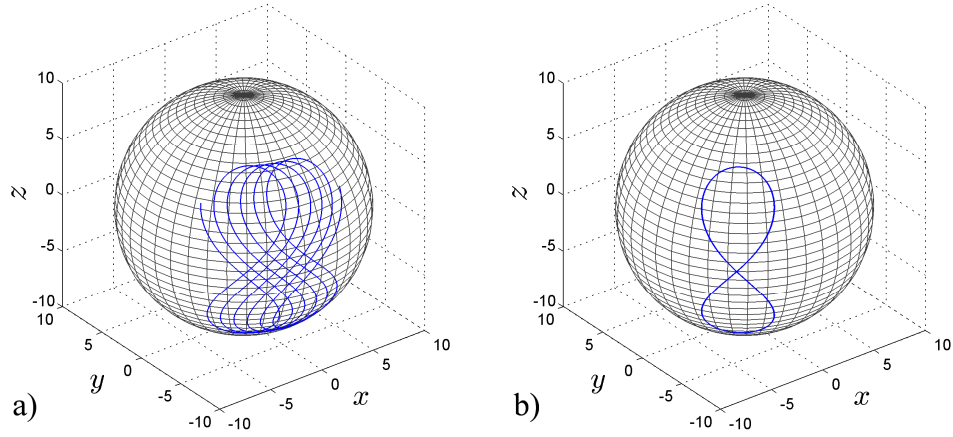

**Figure 13.** Two different paths where we can see loops in the spherical surface. In (a) we have $p_\varphi = -0.71k$ and we have a drift to the east in the azimuthal motion. In (b) we have $p_\varphi = -0.722k$ and there is no drift, so the trajectory in this case is periodic with an oscillation in the azimuthal angle. In both graphs the initial values are $\theta_0 = \frac{\pi}{3}$ and $p_{\theta 0} = 0.2$ (in arbitrary units). For $p_\varphi < -0.722k$ the drift would be to the west.

## 6. Conclusions

In this work we studied the dynamics of a charge constrained in a spherical surface, in the field of a magnetic dipole which is located in the centre of the sphere. We followed a Lagrangian formalism within the non-relativistic classical regime. Although the model can no longer be applied to the study of cosmic radiation, we believe that it may illustrate very clearly the features of this complex magnetic field through a rather simple physical system. We have not found in the literature the analysis of this restricted Stöermer problem. The system has two degrees of freedom and, as the magnetic field of the dipole has an axis of symmetry and the energy is conserved, two constants of motion are readily identified. The nonlinear equations of motion are formally integrable, however we saw that it is not possible to obtain a full analytical solution. We analysed the dynamics in the polar angle variable and found the equivalence of the system with a one dimensional problem with an effective potential. We observed that this effective potential has the form of a double well potential with its barrier in the equator, $\theta = \pi/2$. We discuss different properties of the paths according to this framework. We saw how the maxima and minima which are unstable and stable points respectively, represent horizontal trajectories for which $\theta$ is a constant. We described the turning points in the potential in terms of the energy of the system. Then, from the equation of motion in $\dot\varphi$ we obtain some properties of the path with respect to the azimuthal motion. Here we are predicting the presence of loops in the trajectories, depending on the value of the integration constant given by the conjugate momentum associated with the azimuthal angle. From a numerical integration of the equations of motion, we plotted different graphs to compare them to the analytical results. These graphs include 3D plots of the paths in the spherical surface which illustrate the rich variety of trajectories of a charge in the field of a magnetic dipole.

## 7. Bibliography

[1] Stöermer C 1955 *The Polar Aurora* Oxford University Press UK
[2] Sandoval-Vallarta S M 1961 *Theory of the Geomagnetic Effect of Cosmic Radiation (Handbuch Der Physik XLVI/I )* Springer-Verlag New York, USA pp. 88-129
[3] Poincaré H 1896 Remarques sur une experience de M. Birkland *Comptes Rendus de l´Academie de Sciences* t. 123 Paris pp. 530-533
[4] Dragt A J 2000 Trapped Orbits in a Magnetic Dipole Field *Reviews of Geophysics* **3** nº 2 pp. 255-298
[5] de Alcantara Bonfim O F, Griffiths D J and Hinkley S 2000 Chaotic and Hyperchaotic Motion of a Charged Particle in a Magnetic Dipole Field *Int. J. Bifurcation Chaos* **10** nº 01 p. 265
[6] Jose J V and Saletan E J 2006 *Classical Dynamics: a contemporary approach*, Cambridge University Press New York,USA
[7] Goldstein H 1980 *Classical Mechanics*, Addison-Wesley *2nd* Edition New York USA
[8] Noether E 1918 Invariante Variationsprobleme *Nachr. D.König. Gesellsch. D. Wiss. Zu Götingen, Math-phys. Klasse* pp. 235-257

*Emilio Cortés and D. Cortés-Poza*